\documentclass[a4paper]{EslabStyle}
\usepackage{graphics}

\begin{document}

\title{The Formation History of Globular Clusters}

\author{Dean E. McLaughlin\inst{1,2}}
  \institute{Department of Astronomy, 601 Campbell Hall, University
  of California, Berkeley, CA 94720
  \and Hubble Fellow}

\maketitle 

\begin{abstract}

The properties of old globular cluster systems in galaxy halos are used to
infer quantitative constraints on aspects of star formation
that are arguably as relevant in a present-day context as
they were during the protogalactic epoch. First, the spatial distribution of
globulars in three large galaxies, together with trends in total cluster
population vs.~galaxy luminosity for 97 early-type systems plus the halo of
the Milky Way, imply that bound
stellar clusters formed with an essentially universal efficiency throughout
early protogalaxies: by mass, always $\epsilon_{\rm cl}=0.26\%\pm0.05\%$ of
star-forming gas was converted into globulars rather than halo field stars.
That this fraction is so robust in the face of extreme variations
in local and global galaxy environment suggests that any parcel of gas needs
primarily to exceed a {\it relative} density threshold in order to form a bound
cluster of stars. Second, it is shown that a strict scaling between total
binding energy, luminosity, and Galactocentric position, $E_b=7.2\times10^{39}
\,{\rm erg}\,(L/L_\odot)^{2.05}\,(r_{\rm gc}/8\,{\rm kpc})^{-0.4}$, is a
defining equation for a fundamental plane of Galactic globular clusters. The
characteristics of this plane, which subsumes {\it all other} observable
correlations between the structural parameters of globulars, provide a
small but complete set of facts that must be explained by theories of cluster
formation and evolution in the Milky Way.
It is suggested that the
$E_b(L,r_{\rm gc})$ relation specifically resulted from star formation
efficiencies having been systematically higher inside more massive
protoglobular gas clumps.

\end{abstract}

\section{Introduction}
\label{sec:1}

The old globular cluster systems (GCSs) in galaxy halos are of especial
interest in a conference such as this, as they stand potentially to 
shed light on both local and global star formation in both
protogalactic and present-day settings. Because of their great age, their
ubiquity in galaxies of most any Hubble type, and their impressive spatial
extent (globulars can be found as far as 100 kpc away from the centers of
large galaxies), the integrated properties of GCSs clearly had to have been
influenced by very general, large-scale star-formation processes in the
early universe. But at the same time, most new stars today, whether in
the Galactic disk or in galaxy mergers and starbursts, are born not in
isolation
but in groups. To be sure, the formation of a bona fide star cluster is a
{\it rare} event (only a very small fraction of young stellar groups emerge
from their natal clumps of gas as gravitationally bound units) but, as will be
argued here, it is one that occurs {\it regularly}. Thus, individual globular
clusters must also be viewed---even if in a limit---as the products of a
robust mode of smaller-scale star formation that has always been viable.

It should come as no surprise, then, that there is no definitive
theoretical description of the formation history of globular clusters;
while some concepts have been identified that do seem likely to survive as
elements of a correct theory in the future, at this point there is simply no
model that can claim completeness. Detailed discussions of the many theories
already in the literature may be found in, e.g., \cite*{ash98} or
\cite*{mey97}. The focus here will instead be on recent progress in
extracting quantitative and (as nearly as possible) model-independent
constraints from the data on many GCSs. A similarly empirical discussion is
given by \cite*{har00}, with an eye mostly to implications for large-scale
aspects of galaxy formation and evolution (see also \cite{ash98}). In what
follows, particular emphasis will be placed on applications to the
problem of {\it generic} star formation on subgalactic scales.

A serious concern, when trying to use GCSs in this way, is the influence of
dynamical evolution on the gross properties of cluster systems that
have been immersed for a Hubble time in the tidal fields of their parent
galaxies. Two-body relaxation and evaporation, disk- and bulge-shock heating, 
chaotic scattering or disruption by a compact nucleus, and
dynamical friction: all of these processes whittle away at clusters
individually and collectively (\cite{spi87}; \cite{agu88}; \cite{ost89};
\cite{cap93}; \cite{mwa97}; \cite{mwb97}; \cite{mwc97}; \cite{ves97}). In the
case of the Milky Way particularly, the net effect is to define a roughly
triangular region in mass--radius space, within which globulars are predicted
to survive a 10-Gyr dynamical onslaught (\cite{fal77})---the implication
being that any clusters born outside such a ``survival triangle'' would have
disappeared, taking with them vital information on their birth properties.
However, further investigation shows that Milky Way clusters located at large
Galactocentric radii ($r_{\rm gc}\ga 3$ kpc, roughly the effective radius of
the bulge) {\it do not fill} their expected survival triangles (\cite{cap84}; 
\cite{gne97}). Similarly, evolutionary calculations geared to conditions in
the giant elliptical M87 (\cite{mwb97}) suggest that the damage done to an
initial GCS may be largely confined to within an effective radius in that
galaxy as well. Thus, there are some features of globular clusters, and of
GCSs, that are {\it not} dictated purely by evolution from some completely
obscured initial conditions, and while care must be taken in identifying such
observables, the task is not an impossible one.

Following the suggestion that large fractions of the GCSs in ellipticals may
have formed in major mergers (e.g., \cite{sch87}; \cite{ash92}), and the
related discovery of young, massive star clusters in systems like the Antennae
galaxies (\cite{whi95}), much recent discussion in this field has centered on
the interpretation of (often) bi- and multimodal distributions of globular
cluster colors (as indicators of integrated metallicity) in relation to their
host galaxies' dynamical histories (e.g., \cite{zep93}; \cite{for97};
\cite{cot98}; \cite{kis98}; \cite{kig99}; \cite{cot00}).
However, as will be discussed further in \S\ref{sec:3}, metallicity is
completely decoupled from the other basic properties of
individual globulars in the Milky Way; thus, while GCS colors are of interest
in the context of galaxy formation and chemical evolution, they seem to be of
marginal importance to the star-formation process itself. In this connection,
it is worth noting explicitly that literally thousands of genuinely old
globulars have now been identified, in many galaxies besides our own, with
colors as red as or redder than solar: {\it high metallicity has never posed
any apparent obstacle to the formation of massive star clusters}. This
cautions against theories of globular formation such as those of
\cite*{fal85} and \cite*{mur93}, which, at least in their current form, are
able to account only for metal-poor objects.

Of rather more direct relevance to the small-scale problem are the
luminosity or mass functions of cluster systems. It is well known that the
overall range of Galactic globular cluster masses, $m\sim 10^4$--$10^6\
M_\odot$, and the mean value $\langle m\rangle=2.4\times10^5\ M_\odot$,
are essentially universal properties of other GCSs.
Traditionally, theories of globular cluster formation attempted to explain
just this basic mass scale (e.g., \cite{pee68}; \cite{fal85}).
More recently, however, attention has turned to the full {\it
mass spectrum}, $d{\cal N}/dm$ (the number of clusters with mass $m$ in a
single galaxy), which---in the regime usually observed, $m\ga 10^5\ 
M_\odot$---is similar (though not identical) from galaxy to galaxy; shows no
detectable variations with radius in any one system; and is remarkably similar
to the mass function, $d{\cal N}/dm\propto m^{-1.7}$ or so, of the dense gas
clumps currently forming stars in the giant molecular clouds of the Milky Way
disk (\cite{har94}). These facts have suggested a general physical picture,
developed by \cite*{har94} building on earlier arguments by
\cite*{lar88} and \cite*{lar93}, in which handfuls of globulars form in each
of many protogalactic fragments whose gas masses (of order $10^9\ M_\odot$),
sizes, and internal velocity dispersions correspond to disk GMCs scaled up by
some three orders of magnitude in mass. Such a picture has the obvious appeal
of being at least conceptually consistent both with current models of
hierarchical galaxy formation (there is clearly some affinity with the classic
picture of \cite{sea78} as well) and with the observed pattern of present-day
star formation. It also has more specific attractions (e.g., the
protogalactic fragments and their protoglobulars are presumed to be supported
largely by nonthermal mechanisms, thus allowing for metal-rich protoglobulars
with the correct mass scale) but many details remain to be worked out.
See \cite*{mcl96} (and compare the rather different view of \cite{elm97}) for
an attempt at a quantitative theory for GCS mass spectra which, with the
\cite*{har94} framework as a backdrop, makes explicit use of the emerging links
between present-day and protogalactic star formation.

In addition to color and luminosity distributions, photometric studies of
extragalactic GCSs yield estimates of their specific frequencies (total
cluster populations, normalized to the parent galaxy luminosity) and spatial
structures (number density of globulars as a function of galactocentric
position). These are the focus of \S\ref{sec:2}, where it is shown that a large
sample of early-type galaxies display a common ratio of GCS mass to
total mass in stars and (hot) gas---an apparently universal efficiency for the
formation of globular clusters (\cite{mcl99}). Finally, \S\ref{sec:3}
discusses the binding energies of
globular clusters in the Milky Way (\cite{mcl00}). It is
shown that a strong correlation between binding energy, total cluster
luminosity, and Galactocentric position is instrumental in defining a
fundamental plane for Galactic globulars---which, by incorporating every
one of their many other structural correlations, systematically reduces these
data to the smallest possible set of {\it independent} physical constraints
for theories of cluster formation and evolution.

\section{The Efficiency of Cluster Formation}
\label{sec:2}

The possibility that globular cluster systems can be connected not only to
galaxy formation, but to ongoing star formation as well, is suggested by the
fact that this latter process operates largely in a {\it clustered
mode}. One dramatic example of this is the situation in the Orion B (L1630)
molecular cloud, where 96\% of a complete sample of young stellar
objects are physically associated with just four dense clumps of gas, each
containing $>300\,M_\odot$ of material (\cite{lad91}; \cite{lad92}). More
generally, \cite*{pat94} note that this is just a result of the
different power-law slopes in the mass function of molecular clumps (as above,
$d{\cal N}/dm\propto m^{-1.7}$, so that the largest clumps, which always weigh
in at $10^2$--$10^3\,M_\odot$, contain  most of the star-forming gas mass in
any molecular cloud) and the stellar initial mass function ($d{\cal N}/dm
\propto m^{-2.35}$, putting most of the mass in young stars into $\la1\,
M_\odot$ objects). In more extreme environments, the ``super''
star clusters---luminous, blue, compact associations with integrated properties
roughly consistent with those expected of young globulars---found in many
merging and starburst galaxies can account for as much as $\sim$20\% of the
UV light from such systems (\cite{meu95}).

Again, however, this is not to say that all, or even most, stars are born
into true clusters that exist for any length of time as systems with negative
energy. At some point during the collapse and fragmentation of a cluster-sized
cloud of gas, the massive stars which it has formed will expel any remaining
gas by the combined action of their stellar winds, photoionization, and
supernova explosions. If the {\it cumulative} star formation efficiency (SFE)
of the cloud, $\eta\equiv M_{\rm stars}/(M_{\rm stars}+M_{\rm gas})$, is below
a critical threshold when the gas is lost, then the blow-out removes
sufficient energy that the stellar group left behind is unbound and disperses
into the field. The precise value of this threshold depends on details of the
dynamics and magnetic field in the gas cloud before its self-destruction, and
on the timescale over which the massive stars dispel the gas; but various
estimates place it in the range $\eta_{\rm crit}\sim 0.2$--0.5 (e.g.,
\cite{hil80}; \cite{mat83}; \cite{elm85}; \cite{ver90}; and see \cite{goo97}
for a discussion of globulars specifically).

A general theory of star formation must therefore be able to anticipate the
final cumulative SFE in any single piece of gas with (say) a given mass and
density, and thereby predict whether or not it will form a bound cluster. No
such theory yet exists. It is possible, however, to empirically estimate the
{\it probability} that a cluster-sized cloud of gas is able to achieve
$\eta>\eta_{\rm crit}$. This probability---or, equivalently, that fraction of
an ensemble of massive star-forming clouds which manages to produce bound
stellar systems---is referred to here as the efficiency of cluster formation.
To get a handle on this for globulars in particular, \cite*{mcl99} works in
terms of the {\it mass} fraction

\begin{equation}
\hfil
\epsilon_{\rm cl}\equiv M_{\rm gcs}^{\rm init}/M_{\rm gas}^{\rm init}\ \ ,
\hfil
\label{eq:1}
\end{equation}

\noindent where $M_{\rm gas}^{\rm init}$ refers to the total gas supply that
was available to form stars in a protogalaxy---whether in a monolithic
collapse or a slower assembly of many distinct, subgalactic clumps is
unimportant---and $M_{\rm gcs}^{\rm init}$ is the total mass of all globulars
formed in that gas. The advantage of this definition for $\epsilon_{\rm cl}$
is that the total mass of a GCS is expected to be very well preserved over
the course of dynamical evolution in a galaxy; presently observed values are
reasonable indicators of the initial quantities. This is ultimately due to the
fact that GCS mass spectra are shallow enough that (again, like the clumps in
Galactic molecular clouds) most of any one system's mass is contained in its
most massive clusters. But most of the destruction processes mentioned in
\S\ref{sec:1} operate most effectively against {\it low}-mass globulars, which
may be lost in great numbers (substantially affecting the total GCS population,
${\cal N}_{\rm tot}$) while decreasing the integrated $M_{\rm gcs}$ by
as little as $\sim$25\% (\cite{mcl99}).

Until very recently, it was generally assumed that a galaxy's total luminosity,
or stellar mass, was an adequate stand-in for $M_{\rm gas}^{\rm init}$. Thus,
the number of globulars per unit of halo light has long been taken as a direct
tracer of the efficiency of globular cluster formation in galaxies. However,
this approach leads quickly to two interesting problems.

\subsection{global and local specific frequencies}
\label{sec:21}

Specific frequency was originally defined by \cite*{hvd81} as a global
property of galaxies. It is nominally the ratio, modulo a convenient
normalization, of the total GCS population to the total $V$-band light
integrated over an entire galaxy:

\begin{equation}
\hfil
S_N\equiv{\cal N}_{\rm tot}\times 10^{0.4(M_V+15)} =
8.55\times10^7\,\left({\cal N}_{\rm tot}/L_{V,{\rm gal}}\right)
\hfil
\label{eq:2}
\end{equation}

\noindent Most subsequent studies of GCSs have therefore estimated their total
populations and cited $S_N$-values according to equation (\ref{eq:2}). It is
more useful, however, following the discussion just above, to refer to total
GCS and stellar masses; thus,

\begin{equation}
\hfil
S_N \simeq 2500
\left({{\langle m\rangle}\over{2.4\times10^5\ M_\odot}}\right)^{-1}
\left({{\Upsilon_{V,{\rm gal}}}\over{7\ M_\odot\,L_\odot^{-1}}}\right)
{{M_{\rm gcs}}\over{M_{\rm stars}}}
\hfil
\label{eq:3}
\end{equation}

\noindent for a standard mean globular cluster mass $\langle m \rangle$ and a
representative stellar mass-to-light ratio, $\Upsilon_{V,{\rm gal}}$,
appropriate to the {\it cores} of large ellipticals (which value is used so as
not to include any nonbaryonic dark matter in the galaxy mass). Note that
$\langle m\rangle$ is not observed to deviate significantly from the Milky Way
value, either from galaxy to galaxy or from place to place within any one
system, but that $\Upsilon_{V,{\rm gal}}$ {\it does} vary systematically, as
a function of luminosity, among large ellipticals (e.g.,
\cite{vdm91}).

\begin{figure}[!b]
\resizebox{\hsize}{2.6truein}
{\includegraphics{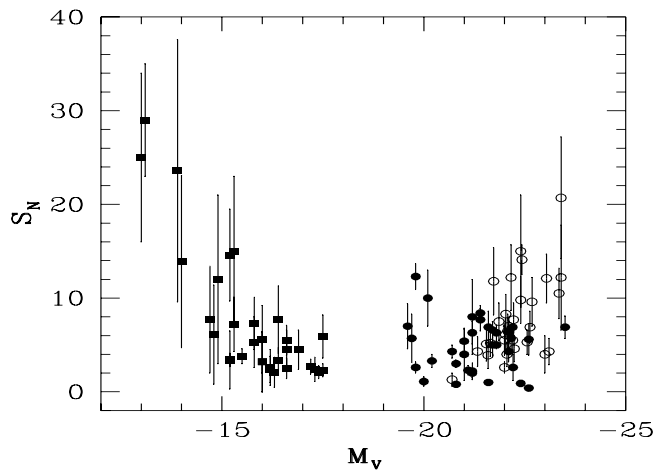}}
\caption{\rm First specific frequency problem in 97 early-type galaxies.
\label{fig1}}
\end{figure}

Global specific frequencies in a large sample of early-type galaxies are
shown in Fig.~\ref{fig1}. (The GCSs of spirals are generally less populous,
and often more difficult to identify, than those in elliptical systems; thus,
the data on late-type galaxies are relatively sparse.) The square points
correspond to dwarf ellipticals and spheroidals, some in the Local Group (the
two faintest objects are the Fornax and Sagittarius dwarfs) and others in the
Virgo cluster (\cite{dur96}; \cite{mil98}); filled circles represent regular
giant ellipticals in a wide range of field and cluster environments (see, e.g.,
\cite{har91} and \cite{kis97}); and open circles stand for the centrally
located galaxies (which often are also the brightest) in a large
number of groups and clusters (\cite{bla97}; \cite{btm97}; \cite{har98}).

Three points are immediately apparent. First, among normal gE's an average
$S_N\approx 5$ (roughly, $M_{\rm gcs}/M_{\rm stars}\sim 0.002$) is
indicated. Second, the specific frequencies of central
galaxies in groups and clusters show a systematic departure from this
``typical'' value, increasing strongly towards brighter galaxy magnitudes.
And third, while the brightest of the dwarf ellipticals have $S_N$ comparable
to the giants, the ratio increases towards {\it fainter} luminosities in these
small systems. All in all, $S_N$ ranges over more than a factor of 20 in
early-type galaxies. If the ratio $M_{\rm gcs}/M_{\rm stars}$ were a good
approximation to $\epsilon_{\rm cl}$ in equation (\ref{eq:1}), the implication
would be that the basic efficiency of cluster formation also varied
drastically---and in a {\it non-monotonic} fashion---from galaxy to galaxy.
This is the first specific frequency problem. Although it has been much
discussed in the literature (see \cite{mcl99}, \cite{elm00}, or
\cite{har00} for recent reviews and references), no satisfactory explanation
(or prediction) of it has ever been advanced.

Very closely related to this, {\it local} specific frequencies may be defined
at different projected radii within a single galaxy, by taking the ratio
of its GCS surface (number) density profile, $N_{\rm cl}(R_{\rm gc})$, and
its $V$-band light intensity, $I_V(R_{\rm gc})$, normalized as in equation
(\ref{eq:2}). (This is then proportional to the ratio of cluster and field-star
{\it mass} densities, just as in eq.~[\ref{eq:3}].) Beyond an effective radius
or so (where the effects of dynamical evolution on the GCS are presumably
minimized), it is found in {\it some} galaxies that local specific frequencies
{\it increase outwards}; in others, they remain constant. Equivalently,
some galaxies' GCSs are significantly more extended than their stellar halos,
while others' GCS and field-star distributions trace each other accurately on
large spatial scales. If the simple assumption $S_N\propto\epsilon_{\rm cl}$
were applied here, it would appear to suggest that bound stellar
clusters were sometimes (but not always, and for reasons completely unknown)
{\it more} likely to form at {\it larger} distances from the centers of
galaxies, in gas that was presumably at lower ambient densities and pressures.
This is the second specific frequency problem.

\begin{figure}[!t]
\resizebox{\hsize}{2.9truein}
{\includegraphics{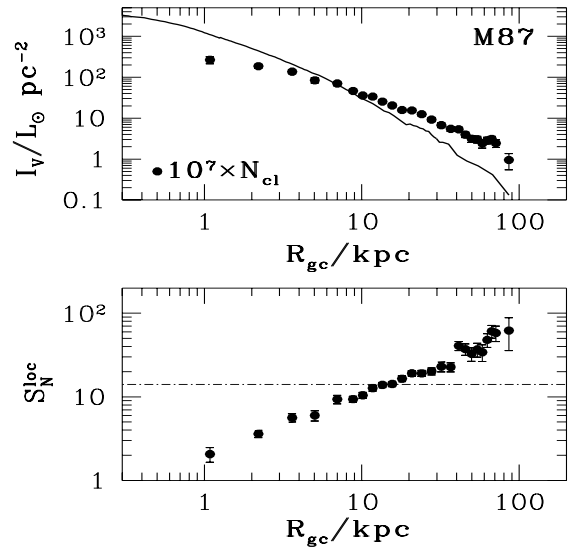}}
\caption{\rm Second specific frequency problem as seen in M87. A distance of
15 Mpc to the galaxy has been assumed.
\label{fig2}}
\end{figure}

Figure \ref{fig2} illustrates this situation in M87, the cD
galaxy at the center of the Virgo Cluster and the first system for which the
effect was shown convincingly to exist (\cite{har86}). The solid line in the
top panel is the galaxy light profile, derived from the surface photometry of
\cite*{dvn78}, and the points trace the projected GCS number
densities (in units of pc$^{-2}$ and scaled up for a direct
comparison with the stellar densities) from the combined data of
\cite*{mcl93}, \cite*{mcl95}, and \cite*{har86} (see \cite{mcl99}). It is
clear that the radial gradient of $N_{\rm cl}$ is significantly shallower than
that of $I_V$, leading directly to the strongly increasing local specific
frequency profile in the bottom panel.

The horizontal line in the bottom of Fig.~\ref{fig2} marks the globally
averaged $S_N$ for M87 as a whole: $14.1\pm1.6$, three times higher than
``normal'' for giant ellipticals (\cite{har98}). That is, M87 also suffers
from the first specific frequency problem. This is one strong hint that the two
$S_N$ problems are really just different aspects of a single basic phenomenon.
What this might be has become clear only with a
homogeneous survey of the central galaxies (simply BCGs hereafter)
in 21 Abell clusters by \cite*{bla97}, \cite*{btm97}, and \cite*{bla99}.

\subsection{x-ray gas and a universal $\epsilon_{\rm cl}$} 
\label{sec:22}

Blakeslee has found (see also \cite{wes95}) that the global specific
frequencies of BCGs increase systematically with the soft X-ray luminosity
(averaged over $\sim$500-kpc scales) of the hot gas in their parent clusters.
He then uses the details of this correlation to argue that the number of
globulars in the cores of galaxy clusters scales in direct proportion to the
total mass there, with one found for every 1--$2\times10^9\,M_\odot$ of
stars, gas, {\it and dark matter}. This implies that the first $S_N$ problem
stems (in BCGs, at least) from a tendency for brighter galaxies
to be ``underluminous,'' for the amount of gas and dark matter associated with
them, rather than overabundant in globular clusters. Strictly from a
star-formation point of view, however, only the baryons are of interest; thus,
\cite*{har98} suggest, in essence, that if the dark matter were left out,
the global mass ratio

\begin{equation}
\hfil
\widehat{\epsilon}_{\rm cl}=M_{\rm gcs}/(M_{\rm gas}+M_{\rm stars})
\hfil
\label{eq:4}
\end{equation}

\noindent might itself be constant, not only among BCGs but in other galaxies
as well.\footnotemark
\footnotetext{As \cite*{bla99} discusses at length, his
ratio of globulars per unit total mass and the ratio of equation
(\ref{eq:4}) can {\it both} be constants in BCGs if the baryon fraction in
the cores of galaxy clusters is also roughly universal (on the order of
10\%). But as reasonable as it may seem, this possibility is in general
unproven, and it is not obvious a priori that the two efficiencies are
necessarily equivalent. Also, the constancy of Blakeslee's ratio has been
demonstrated neither globally for objects other than BCGs nor locally as a
function of position inside any one galaxy. The behavior of $\widehat{\epsilon}
_{\rm cl}$ {\it in general} can therefore not be anticipated from Blakeslee's
work.}
The first specific frequency problem would still arise more or less
as Blakeslee suggested, with the (observed) larger gas fractions
$M_{\rm gas}/M_{\rm stars}$ in brighter galaxies resulting also in
a higher global $S_N\propto M_{\rm gcs}/M_{\rm stars}$. In addition,
\cite*{mcl99} notes that if the local $\widehat{\epsilon}_{\rm cl}$---defined
in the obvious way as a ratio of densities---were also constant
as a function of radius in galaxies, then the fact that the X-ray gas in
ellipticals tends to be hotter and more spatially extended than the stellar
distribution could cause the second specific frequency problem in
gas-rich galaxies that also have a high global $S_N$.

The underlying idea here is, of course, that the present-day sum $(M_{\rm gas}
+ M_{\rm stars})$ should be a better indicator of the ``initial'' gas mass in
large galaxies, and that equation (\ref{eq:4}) may therefore be more accurate
than $S_N$ as an estimate of the true cluster formation efficiency in equation
(\ref{eq:1}). Admittedly, $\widehat{\epsilon}_{\rm cl}$ is still only an
observable {\it approximation} to $\epsilon_{\rm cl}$; indeed, in a
hierarchical universe it may be difficult to say precisely what is meant by
$M_{\rm gas}^{\rm init}$ in the first place. Particularly in a large galaxy
which may have accreted gas (and stars and globulars) over extended lengths of
time, the observable $\widehat{\epsilon}_{\rm cl}$ must be viewed as a
mass-weighted average over a complex evolutionary history including a
potentially large number of galaxy interactions and many discrete
star-formation episodes. However, the current ratio of $M_{\rm gcs}$ to
$(M_{\rm gas}+M_{\rm stars})$ does have an essentially universal value in old
GCSs---including those in regular gE's and early-type dwarfs as well as
BCGs---arguing that it may be a good reflection after all of the real
$\epsilon_{\rm cl}$.

\begin{figure}[!t]
\resizebox{\hsize}{2.9truein}
{\includegraphics{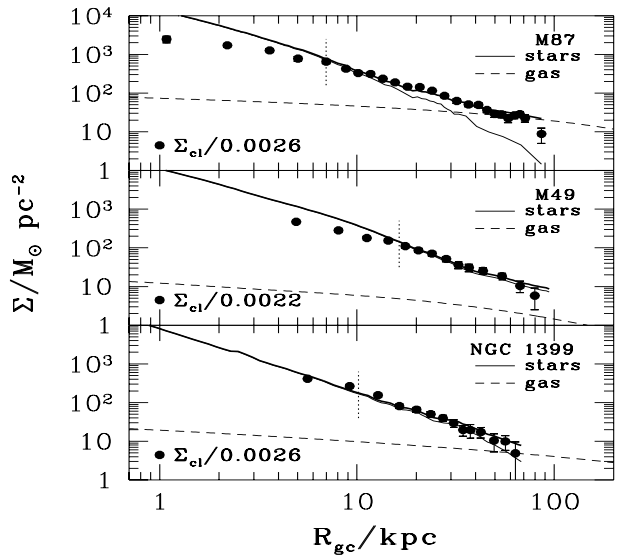}}
\caption{\rm Constant {\it local} cluster formation efficiency in M87, M49,
and NGC 1399 (McLaughlin 1999). Bold lines in all panels trace the {\it sum}
of star and gas surface densities. Broken vertical lines mark the stellar
effective radius of each galaxy.
\label{fig3}}
\end{figure}

Figure \ref{fig3} shows this first for the {\it local} $\widehat{\epsilon}
_{\rm cl}$ in M87 and in the bright ellipticals M49 (also in the Virgo
Cluster) and NGC 1399 (BCG in the Fornax Cluster): The ratio of
projected GCS mass densities ($\Sigma_{\rm cl}\equiv\langle m\rangle\,N_{\rm
cl}$) to the sum of stellar and gas mass densities ($\Upsilon_{V,{\rm gal}}\,
I_V + \Sigma_{\rm gas}$, with $\Upsilon_{V,{\rm gal}}$ measured separately for
each system) is constant beyond an effective radius in the
galaxies.\footnotemark
\footnotetext{The departure of the GCS densities from the stellar profiles at
smaller radii in M87 and M49 is undoubtedly significant, but it is not clear
whether this is due to a real decrease in the true cluster formation
efficiency there, or to dynamical depletion of the initial GCS, or to
substantial dissipation in the gas that formed the field stars (after the
globulars were already in place) in the innermost regions of the galaxies.}
The construction of the individual density profiles is discussed in detail by
\cite*{mcl99}, where a comparison of the deprojected quantities is also made,
confirming the basic result and giving essentially the same numbers
for $\widehat{\epsilon}_{\rm cl}$.
Evidently, including the gas does alleviate the second specific frequency
problem, in just the sense suggested above: $S_N$ increases with radius in M87
because of a locally varying gas-to-star mass ratio, rather than any change in
the fundamental $\widehat{\epsilon}_{\rm cl}$; and in the
comparatively gas-poor M49 and NGC 1399, there is no $S_N$ problem in the
first place. Moreover, the first $S_N$ problem is similarly removed when the
X-ray gas is taken into account: Although the total specific frequencies of
M87, M49, and NGC 1399 are significantly different (at 14.1, 4.7, and
6--7), their local GCS mass ratios at large radii are consistent with a
single value: $\widehat{\epsilon}_{\rm cl}=0.0026\pm0.0005$ in the
mean. Finally, since this is also independent of galactocentric radius, it
is the same as the {\it global} cluster formation efficiency in each case.

\begin{figure}[!t]
\null\vskip-0.2truein
\resizebox{\hsize}{2.9truein}
{\includegraphics{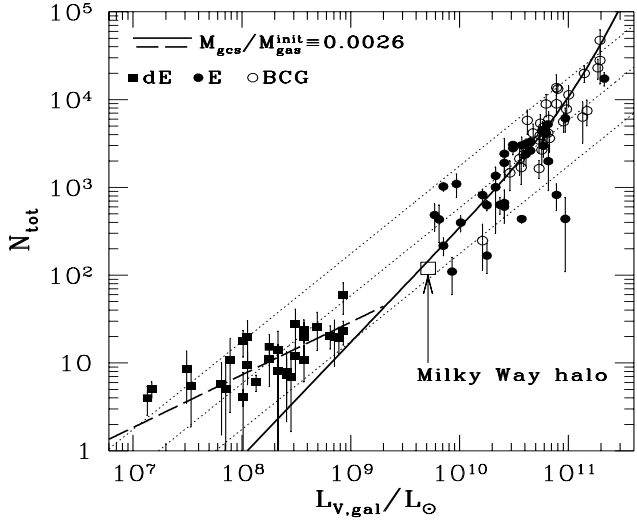}}
\caption{\rm Total GCS populations and galaxy luminosities for the same systems
plotted in Fig.~\ref{fig1}. Bold lines are the relations predicted by a
constant {\it global} cluster formation efficiency of $\epsilon_{\rm cl}\equiv
0.0026$. Light, dotted lines represent constant specific frequencies of
$S_N=15$, 5, and 1.5. From McLaughlin (1999).
\label{fig4}}
\end{figure}

These specific examples clearly are consistent with the notion of a universal
$\epsilon_{\rm cl}$. Figure \ref{fig4}, which essentially re-plots the
$S_N$ data in Fig.~\ref{fig1}, confirms it in detail on a global scale.
To understand the bold curves in this Figure, note that equation (\ref{eq:4})
and the definition ${\cal N}_{\rm tot}=M_{\rm gcs}/(2.4\times10^5\,M_\odot)$
can be written as

\begin{equation}
\hfil
{\cal N}_{\rm tot}=4.17\times10^6\,\widehat{\epsilon}_{\rm cl}\,
\left(1+{{M_{\rm gas}}\over{M_{\rm stars}}}\right)\,
\left({{M_{\rm stars}}\over{10^{12}\,M_\odot}}\right)\ .
\hfil
\label{eq:5}
\end{equation}

\noindent The heavy solid line in Fig.~\ref{fig4} is just this equation, with
(i) $\widehat{\epsilon}_{\rm cl}=0.0026$ fixed; (ii) $V$-band galaxy
luminosities converted to stellar masses according to the relation
$\Upsilon_{V,{\rm gal}}=
6.3\ M_\odot\,L_\odot^{-1}\,(L_{V,{\rm gal}}/10^{11}\,L_\odot)^{0.3}$
(\cite{vdm91}); and (iii) galaxy-wide gas-to-star mass ratios
estimated from a combination of fundamental-plane scalings and
the X-ray--optical
luminosity correlation: $M_{\rm gas}/M_{\rm stars}\approx 0.55  \times
(L_{V,{\rm gal}}/10^{11}\,L_\odot)^{1.5}$ (\cite{mcl99}). The
steep upturn in ${\cal N}_{\rm tot}$ at high $L_{V,{\rm gal}}$ (or the
sharp increase in BCG specific frequency) is thus due to the fast-growing
dominance of gas over stars, in
the face of a {\it constant} GCS mass fraction. Towards lower luminosities,
global gas masses become negligible and $S_N$ decreases steadily because of the
systematic decrease in $\Upsilon_{V,{\rm gal}}$ for fundamental-plane
ellipticals. As a result, at the luminosity of the Milky Way spheroid (disk
excluded), an $\widehat{\epsilon}_{\rm cl}$ of 0.0026 also accounts for the
number of halo (metal-poor) Galactic globulars (open square in
Fig.~\ref{fig4}; see \cite{mcl99}).

For the early-type dwarf galaxies with $L_{V,{\rm gal}}\le 2\times10^9\ 
L_\odot$, the gas-to-star mass ratio in equation (\ref{eq:5}) has a different
meaning: The energy of supernova explosions in a single burst of star
formation in one of these small galaxies may have sufficed to expel all
remaining gas from its dark-matter well, and while any such gas would, of
course, no longer be directly observable, a proper estimate of
$\epsilon_{\rm cl}$ must still account for it. The bold, dashed line in
Fig.~\ref{fig4} is one attempt to do this. It represents equation (\ref{eq:5})
given (i) the relation $(1+M_{\rm gas}^{\rm lost}/M_{\rm stars})\simeq
(L_{V,{\rm gal}}/2\times10^9\,L_\odot)^{-0.4}$, from the {\it theory} of
\cite*{dek86}; (ii) a constant
$\Upsilon_{V,{\rm gal}}=2\ M_\odot\,L_\odot^{-1}$ for the stellar
populations; and (iii) once again, a fixed $\widehat{\epsilon}_{\rm cl}=
0.0026$.

Although scatter remains (at the level of factors of $\sim$2) in the observed
${\cal N}_{\rm tot}$ at any given $L_{V,{\rm gal}}$ in
Fig.~\ref{fig4}, it is important that this is more or less random; the
mean {\it trends} in GCS population as a function of luminosity---the essence
of the first specific frequency problem---can be simply explained if the
efficiency of globular cluster formation were constant to first order. Indeed,
deviations in Fig.~\ref{fig4} may reflect the scatter of individual galaxies
about either the fundamental plane or the $L_X$--$L_B$ correlation used to
derive the bold lines there, rather than any significant variations in
$\epsilon_{\rm cl}$. This requires further study on a case-by-case basis, as
does the situation in spirals other than the Milky Way. Similarly, there is
some indication (e.g., \cite{mac99}) that the simple treatment of galactic
winds (i.e., the model of \cite{dek86}) used to correct for the gas lost from
dwarf galaxies may be inadequate. At this point, however, it is more than
plausible to assert that globular clusters formed in dE's and dSph's as in
larger galaxies, always in the same proportion to the total mass of gas that
was initially on hand.

One important consequence of this is that the efficiency of {\it unclustered}
star formation in protogalaxies could {\it not} have been universal.
In both the faintest dE's and the brightest BCGs, globular clusters
apparently formed in precisely the numbers expected of them, while anomalously
low fractions of the initial gas mass were converted into field stars. In
the case of the dwarfs specifically, if even just the idea of the feedback
correction above is basically correct then all the globulars had to have
formed by the time a galactic wind cleared the remaining gas; but this must
have happened before normal numbers of field stars appeared. Thus,
{\it the gas which formed bound star clusters had to have collapsed more
rapidly than that which produced unbound groups and associations}. This implies
that it was only those pieces of gas which locally exceeded some critical
density that were able to attain the cumulative star formation efficiency of
$\eta\ga20\%$--50\% required to form a bound stellar cluster. In addition to
this, the uniformity of $\widehat{\epsilon}_{\rm cl}$
argues---applying as it does over large ranges of radius inside M87, M49, and
NGC 1399, and from dwarfs in the field to BCGs in the cores
of Abell clusters---that the probability of realizing such a high SFE
depended very weakly, if at all, on local or global protogalactic
environment. Quantitative theories of cluster formation should therefore seek
to identify a {\it threshold in relative density}, $\delta\rho/\rho$, that
is always exceeded by $\simeq$0.26\% of the mass fluctuations in any
large body of star-forming gas.

The ``relative'' aspect of such a criterion is crucial; the GCS data militate
strongly against any model relying on parameters that are too sensitive to
environment. One such example is the scenario of \cite*{elm97}, in which
the pressure exerted by a diffuse medium surrounding a dense clump
of gas must exceed a fixed, absolute value in order to produce a
high local $\eta$ and a bound stellar cluster. However, since pressures vary
by orders of magnitude in going from dE's to BCGs, or from large to small
radii in any one galaxy, this idea seems to imply {\it systematic} variations
in $\epsilon_{\rm cl}$ that are not observed.

BCGs present a complex problem in larger-scale galaxy formation, but it is
worth noting that a feedback argument like that applied to dwarfs may also be
relevant to central cluster galaxies like M87 (cf.~\cite{har98}; \cite{mcl99}).
That is, globulars likely also appeared quickly, and in normal numbers,
in the densest of star-forming clumps (perhaps embedded in dwarf-sized
fragments) in these deep potential wells. The gas more slowly forming field
stars could have been virialized thereafter, or moved outwards in slow,
partial galactic winds. The unused gas in this case would have to remain hot
to the present day, and more or less in the vicinity of the parent galaxy,
in order to appear as the X-ray emitting gas that makes $\widehat{\epsilon}_
{\rm cl}$ so constant in Fig.~\ref{fig4}; but this requirement is certainly
consistent with the BCGs being at the centers of clusters. In addition, the
feedback in this scenario would have more effectively truncated the star
formation in the lower-density environs at larger galactocentric radii in
these very large systems, thus giving rise to the second specific frequency
problem as well. There are other possibilities for BCGs, however. It is
conceivable, for instance, that their ``excess'' gas and globulars were both
produced elsewhere in galaxy clusters (in failed dwarfs?) and fell together
onto the central galaxies over a long period of time. These questions need to
be examined in much more detail.

Finally, \cite*{mcl99} argues that the current efficiency of {\it open}
cluster formation in the Galactic disk is also $\sim$0.2--0.4\% by mass. This
figure is much more uncertain than it is in GCSs, and it is essentially an
{\it instantaneous} variant of the time-averaged quantity measured for globular
clusters. Nevertheless, it clearly suggests that whatever quantitative
criterion is ultimately required to explain $\epsilon_{\rm cl}=0.26\%$ in
GCSs may very well prove to be of much wider applicability. (One exception
{\it may} be the formation of massive clusters in mergers and starbursts,
where it has been suggested that $\epsilon_{\rm cl}\sim1$--10\% [e.g.,
\cite{zep99}; \cite{sch99}]. However, this conclusion is very uncertain and
requires more careful investigation.)

\section{Globular Cluster Binding Energies}
\label{sec:3}

The focus to this point has been on the {\it frequency} with which
$\sim$$10^5$--$10^6\,M_\odot$ clumps of gas were able to form stars with a
cumulative efficiency $\eta$ high enough to produce a bound globular cluster.
The impressive regularity of this occurrence is clearly important, as has just
been discussed, and its rarity is significant as well: the small value of
$\epsilon_{\rm cl}=0.26\%$ implies that the local SFE
in an {\it average} bit of protogalactic gas was much lower than
$\eta_{\rm crit}\sim0.2$--0.5 (a fact which is also true of molecular gas in
the Galaxy today). However, these results say nothing of {\it how} an extreme
$\eta$ comes about in any individual gas clump. This is another open
problem in star formation generally. Its solution requires both an
understanding of local star formation laws ($d{\rho}_*/dt$ as a function of
$\rho_{\rm gas}$) and a self-consistent treatment of feedback on
small ($\sim$10--100 pc) scales.

The whole issue is essentially one of energetics in a compact,
gravitationally bound association of gas and embedded young stars: When does
the combined energy injected by all the massive stars equal the binding energy
of whatever gas remains? This point of equality and the corresponding $\eta$
can in principle be identified for any given star formation law and a set of
initial conditions in the original gas. The difficulty lies, of course, in
deciphering what these are; but once this is done, progress will also have been made
in understanding the probability of obtaining $\eta > \eta_{\rm crit}$, i.e.,
the overall efficiency of cluster formation in \S\ref{sec:2}.
One way to begin addressing this complex set of problems empirically is to
compare the final binding energies of stellar clusters with the initial
energies of their gaseous progenitors. This is a straightforward exercise for
the ensemble of globular clusters in the Milky Way.

\cite*{sai79a} evaluated the binding energies $E_b$ for about 10 bright
Galactic globulars, along with a number of dwarf and giant ellipticals. He
claimed that $E_b\propto M^{1.5}$ for gE's and globulars alike, while the
dwarfs fell systematically below this relation (a fact which he subsequently
attributed to the effects of large-scale feedback such as discussed above
[\cite{sai79b}; but cf.~\cite{ben92}]). Twenty years later, the data
required for a calculation of binding energy are available for many more
than ten globular clusters, and they are of higher quality than those
available to Saito.

\cite*{mcl00} computes the binding energies of 109 ``regular'' Galactic
globulars and 30 objects with post--core-collapse (PCC) morphologies. The main
assumption is that single-mass, isotropic \cite*{kin66} models provide a
complete description of the clusters' internal structures. Within this
framework, the fundamental definition $E_b\equiv -(1/2)\int_0^{r_t}
4\pi r^2\rho\phi\,dr$ (with $r_t$ the tidal radius of the cluster) may be
written as

\begin{equation}
\hfil
E_b=1.663\times10^{41}\,{\rm erg}\,
\left({{r_0}\over{{\rm pc}}}\right)^5
\left({{\Upsilon_{V,0}\,j_0}\over{M_\odot\,{\rm pc}^{-3}}}\right)^2
{\cal E}(c)\ \ ,
\hfil
\label{eq:6}
\end{equation}

\noindent where $r_0$ is the model scale radius (\cite{kin66})
$\Upsilon_{V,0}$ is the core mass-to-light ratio; $j_0$ is the central $V$-band
luminosity density; and ${\cal E}(c)$ is a well defined, nonlinear function of
$c\equiv\log\,(r_t/r_0)$, obtained by numerically integrating King models
(\cite{mcl00}).

\begin{figure}[!hb]
\null\vskip-0.46truein
\resizebox{\hsize}{2.72truein}
{\includegraphics{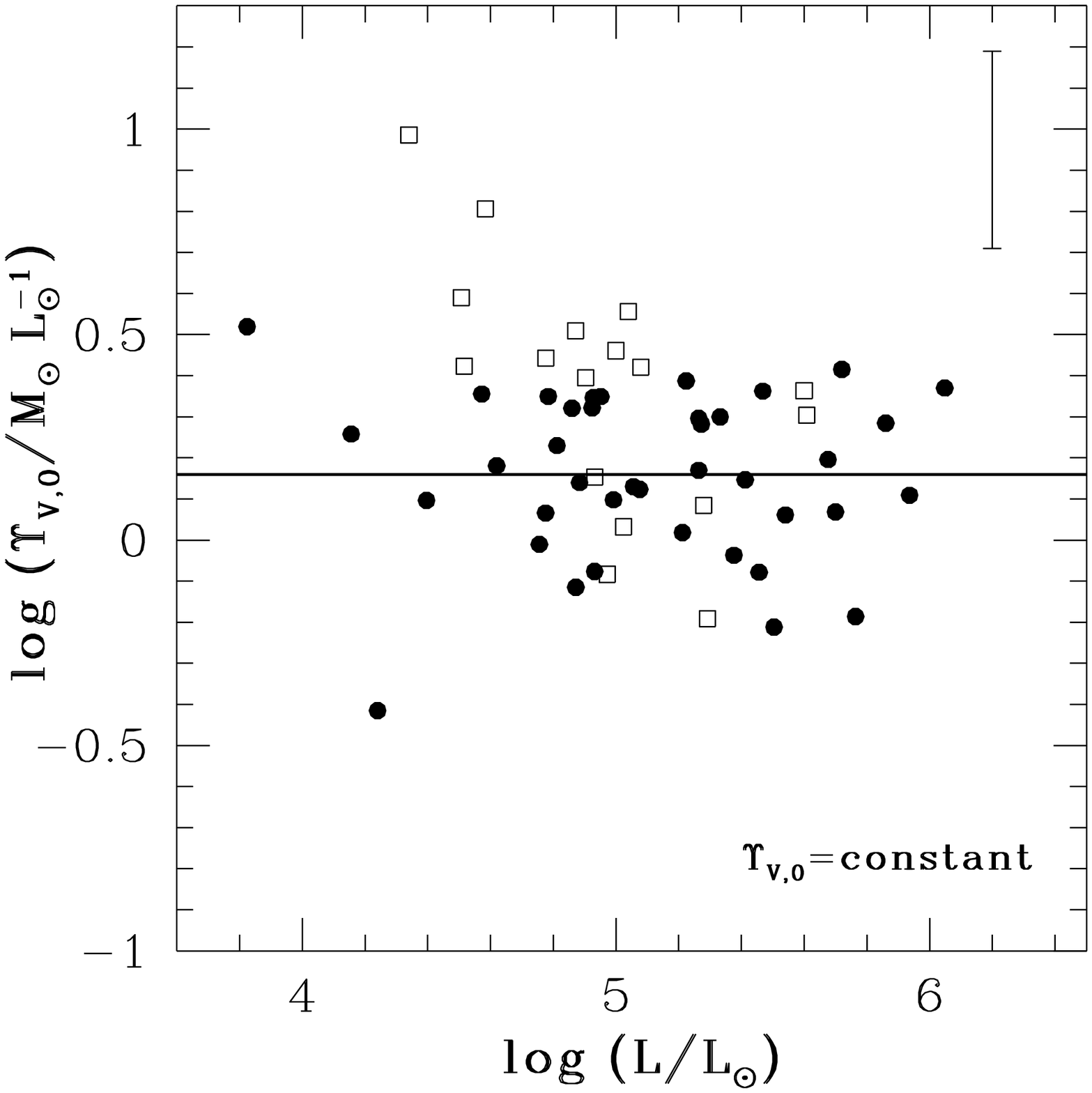}}
\resizebox{\hsize}{2.72truein}
{\includegraphics{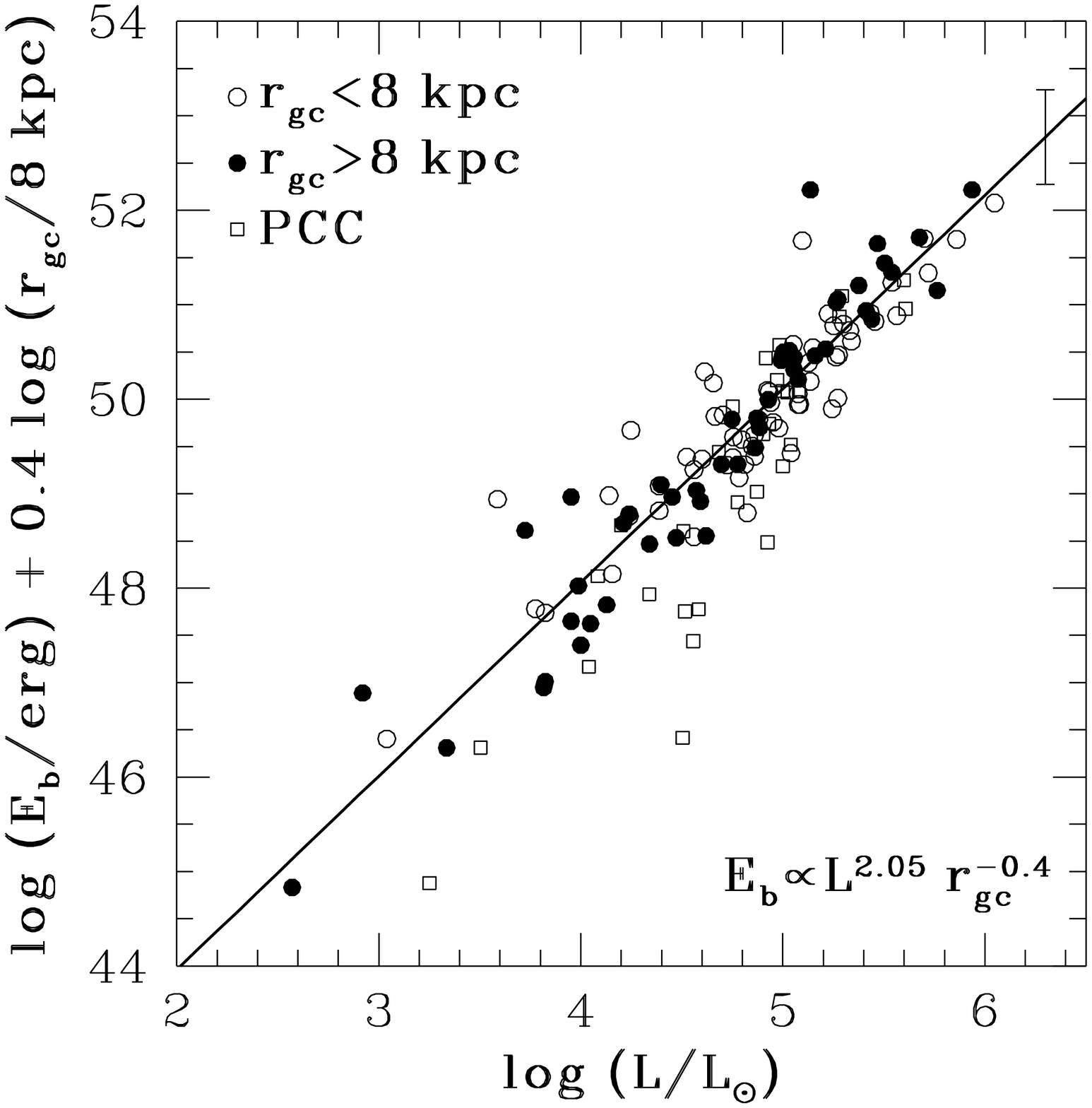}}
\resizebox{\hsize}{2.72truein}
{\includegraphics{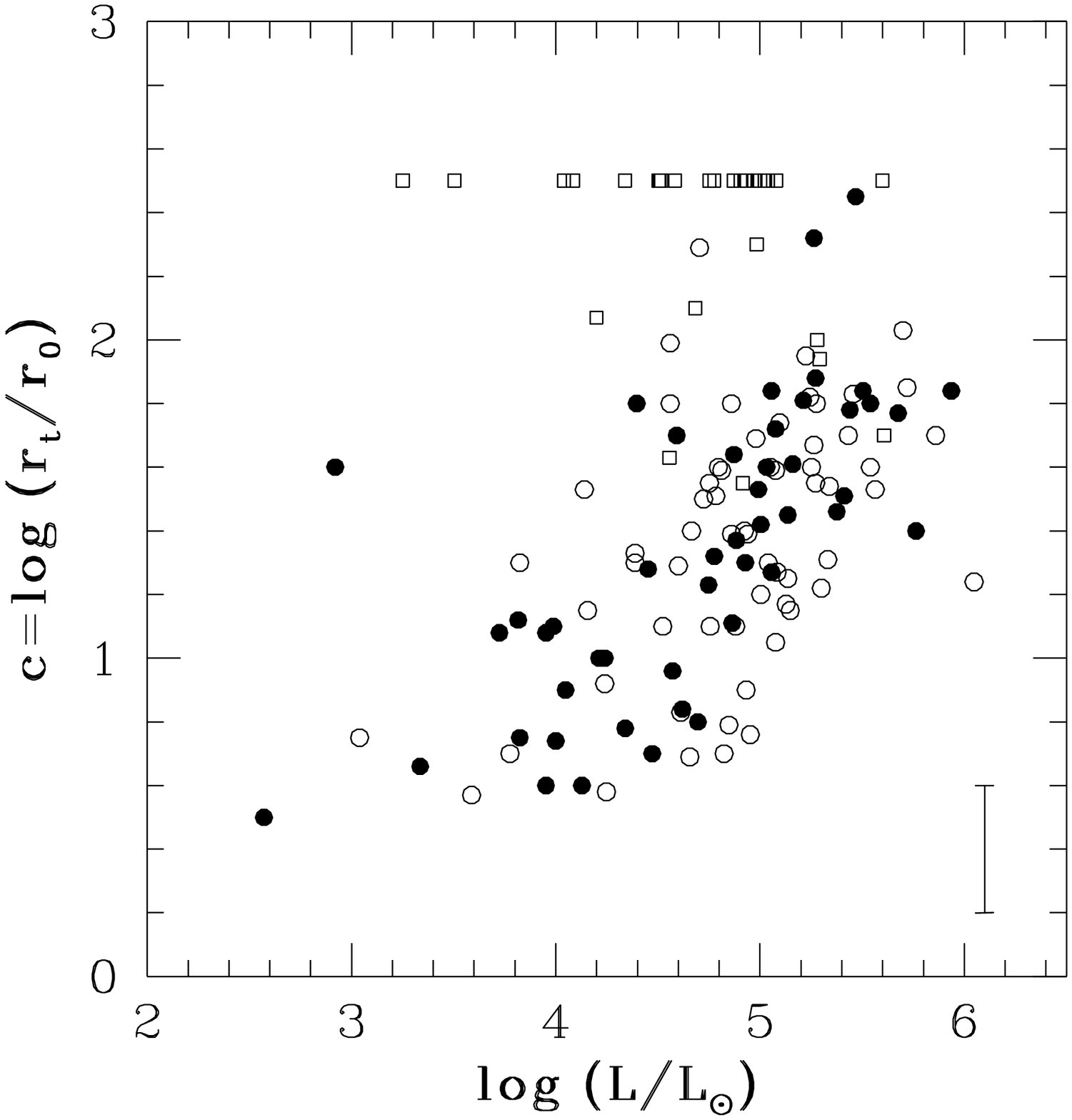}}
\caption{\rm The fundamental plane of Galactic globular clusters (after
McLaughlin 2000). Top panels are two edge-on views; bottom is nearly the
face-on view. All correlations between any other combinations of cluster
observables follow directly from these three relations between
$\Upsilon_{V,0}$, $E_b$, $c$, $L$, and $r_{\rm gc}$.
\label{fig5}}
\end{figure}

Values of $r_0$, $j_0$, and $c$ are given for all Milky Way clusters in the
catalogue of \cite*{har96}. However, a determination of $\Upsilon_{V,0}$
requires measurements of velocity dispersions $\sigma_0$, and these are
available for only a third of the sample, in the compilation of \cite*{pry93}.
For these objects, application of the King-model relation $\Upsilon_{V,0}=
9\sigma_0^2/(4\pi G\,r_0^2\,j_0)$ gives the results shown in the top panel of
Fig.~\ref{fig5}. The regular globulars there (the 39 filled circles) share
a single, {\it constant} core mass-to-light ratio: $\langle \log\,
\Upsilon_{V,0} \rangle = 0.16\pm0.03$ in the mean, and the r.m.s.~scatter
about this is less than the 1-$\sigma$ observational errorbar shown for
$\log\,\Upsilon_{V,0}$. (The results for 17 PCC clusters, plotted as open
squares, are almost certainly spurious [see \cite{mcl00}]. They are shown for
completeness but not included in any quantitative analyses here.) This is
consistent with separate work by \cite*{man91} and \cite*{pry93}.

It can safely be assumed that this same $\Upsilon_{V,0}$ applies to all other
(non-PCC) Galactic globulars, so that $E_b$ can be computed from equation
(\ref{eq:6}) given only $r_0$, $j_0$, and $c$, i.e., on the basis of cluster
photometry or star-count data alone. If this is done for the full \cite*{har96}
catalogue, a very tight correlation between binding energy, total cluster
luminosity, and Galactocentric position is found:

\begin{equation}
\hfil\quad
E_b=7.2\times10^{39}\,{\rm erg}\ (L/L_\odot)^{2.05}\,(r_{\rm gc}/
8\,{\rm kpc})^{-0.4}\ ,
\hfil
\label{eq:7}
\end{equation}

\noindent with uncertainties of about $\pm$0.1 in the fitted powers on $L$ and
$r_{\rm gc}$. This relation is drawn as the line in the middle panel of
Fig.~\ref{fig5}. The r.m.s.~scatter of the regular-cluster data
(filled and open circles) about it is no larger than the typical 1-$\sigma$
observational uncertainty on $\log\,E_b$.

So far as current data can tell, the constancy of $\Upsilon_{V,0}$ and the
scaling of $E_b$ with $L$ and $r_{\rm gc}$ are essentially perfect. Now, in
the context of \cite*{kin66} models, any globular cluster is fully defined
by the specification of just four (nominally) independent physical quantities.
Given the results just presented, it is natural to choose these to be
$\log\,\Upsilon_{V,0}$, $\log\,E_b$, the total $\log\,L$, and the concentration
parameter $c=\log\,(r_t/r_0)$. (Additional factors such as Galactocentric
position or cluster metallicity are quite separate from the model
characterization of a cluster, and they are to be viewed as external
parameters.) But the tight empirical constraints on $\Upsilon_{V,0}$ and $E_b$
mean that they are not actually ``free'' in any real sense; in practice, 
Galactic globulars are only a {\it two-parameter} family, with all internal
properties set by $\log\,L$ and $c$. Equivalently, the clusters are confined
to a {\it fundamental plane} in the larger, four-dimensional space of King
models available to them in principle. The top plots in Fig.~\ref{fig5} are
then just two edge-on views of this plane. Its properties are discussed in
detail by \cite*{mcl00}, where this physical approach to it is also compared
to the more statistical tack taken by \cite*{djo95}, who first claimed its
existence, and to the different interpretation suggested by \cite*{bel98}.

The bottom panel shows the third plot possible in the physical cluster
``basis'' chosen here: concentration vs.~total luminosity. (This is
close to, but not quite, a face-on view of the fundamental plane; see
\cite{mcl00}.) Although it is not one-to-one like those in
the top panels, there is clearly a dependence of $c$ on $\log\,L$ (see also
\cite{vdb94} or \cite{bel96}): roughly, $c\approx -0.55+0.4\,\log\,L$, but the
scatter about this line exceeds the observational errorbar on $c$. Neither
the slope nor the normalization of this rough correlation changes with
Galactocentric position, i.e., the distribution of globulars {\it on} the
fundamental plane is independent of $r_{\rm gc}$.

The mean core mass-to-light ratio is also independent of Galactocentric
radius, and {\it none of the distributions in Fig.~\ref{fig5} depend on
cluster metallicity}. Moreover, since any other property of a
King-model cluster is known once values for $\Upsilon_{V,0}$, $E_b$, $c$,
and $L$ are given, it follows that {\it all} interdependences between {\it any}
globular cluster observables (and there are many; see, e.g., \cite{djo94}) are
perforce equivalent to a combination of (i) a constant $\Upsilon_{V,0}=1.45\
M_\odot\,L_\odot^{-1}$; (ii) equation (\ref{eq:7}) for $E_b$ as a function of
$L$ and $r_{\rm gc}$; (iii) the rough
increase of $c$ with $L$; and (iv) generic King-model definitions.
\cite*{mcl00} derives a complete set of
structural and dynamical correlations to confirm this basic point: {\it only}
the quantitative details of Fig.~\ref{fig5}---and their insensitivity to
metallicity---need be explained in any theory of globular cluster formation
and evolution in our Galaxy.

\begin{figure}[!b]
\resizebox{\hsize}{2.9truein}
{\includegraphics{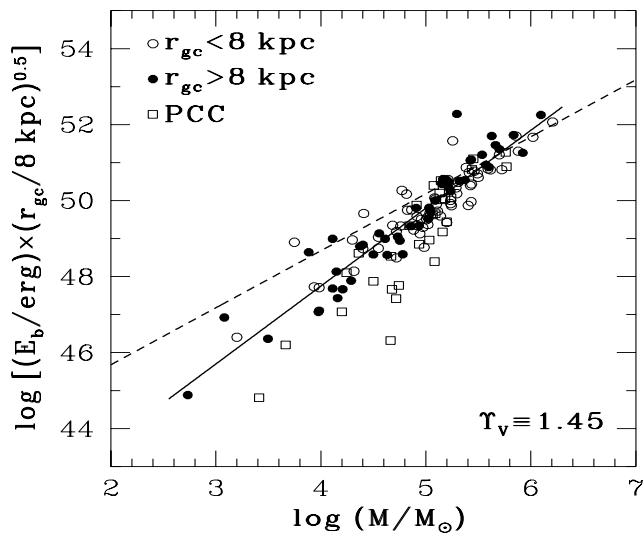}}
\caption{\rm Binding energy vs.~mass for globulars (points; solid
line) and their gaseous progenitors (broken line) in the Galaxy.
Total cluster luminosities are converted to masses by applying the constant
mass-to-light ratio indicated.
\label{fig6}}
\end{figure}

It is then important that the $E_b(L,r_{\rm gc})$ and
$c(L)$ correlations are stronger among clusters outside the Solar circle
(filled circles in the plots) than among those within it (open circles).
Given the relative weakness of dynamical evolution at such large $r_{\rm gc}$,
this is one indication that these fundamental properties of the Galactic
GCS were set largely by the cluster {\it formation} process (see also
\cite{mur92}; \cite{bel96}; \cite{ves97}).

Figure \ref{fig6} finally compares the globular cluster energies to estimates
for the initial values in their progenitors. This is done for the specific
model of \cite*{har94}, in which protoglobular clusters are embedded in
larger protogalactic fragments and have properties analogous to those of the
dense clumps inside present-day molecular clouds (see \S\ref{sec:1}). In
particular, the column densities of the protoclusters are postulated to be
independent of mass but decreasing with Galactocentric radius: $M/\pi R^2
\simeq10^3\
M_\odot$ ${\rm pc}^{-2}\,(r_{\rm gc}/8\,{\rm kpc})^{-1}$, which follows
from the clumps being in hydrostatic equilibrium and from their parent clouds
being themselves surrounded in a diffuse medium virialized in a ``background''
isothermal potential well with a circular velocity of 220 km s$^{-1}$.
This relation then implies $E_b\equiv GM^2/R = 4.8\times10^{42}\ 
{\rm erg}\,(M/M_\odot)^{1.5}\,(r_{\rm gc}/8\,{\rm kpc})^{-0.5}$, which is
drawn as the broken line in Fig.~\ref{fig6}. {\it By construction, this is
precisely the mass-energy relation obeyed today by the massive clumps in
molecular clouds in the Solar neighborhood.} Intriguingly, it is also
the $M$--$E_b$ scaling originally claimed by \cite*{sai79a} for giant
elliptical galaxies and (bright) globular clusters.

The dependence of $E_b$ on $r_{\rm gc}$ in such protoclusters is nearly the
same as that actually found for the globulars today. (It similarly accounts
for the observed increase of cluster radii with $r_{\rm gc}$
[\cite{har94}; cf.~\cite{mur92}]---a trend which is, in fact, equivalent to
the behavior of $E_b$ in eq.~[\ref{eq:7}].) In Fig.~\ref{fig6}, the two are
taken for convenience to be identical, so that the comparison between model
and observed binding energies there is valid at any given Galactocentric
position.

The difference in the {\it slopes} of the two $E_b(M)$ relations is
significant: The ratio of the initial energy of a gaseous clump to the
final $E_b$ of a stellar cluster is unavoidably a function of its cumulative
star formation efficiency $\eta$; but Fig.~\ref{fig6} shows that this ratio of
energies changes systematically as a function of mass, and thus that $\eta$
varied as well. Moreover, the fact that the difference between initial and
final $E_b$ is largest at the lowest masses implies that {\it $\eta$ had to
have been lower in lower-mass protoglobulars}. The details of this behavior
must rely on the density and velocity structure of the initial gas; the
timescale over which feedback expels unused gas; re-expansion of the stars
after such gas loss; and other such specifics which are model-dependent to
some extent. The inference on the qualitative behavior of $\eta$ as a function
of protocluster gas mass is, however, robust.

A more quantitative discussion of Fig.~\ref{fig6}---including its implications
for the mass function of GCSs, which will differ from the mass functions of
gaseous protoclusters if $\eta$ varied systematically from one to the 
other---has to be deferred (McLaughlin, in progress). But this evidence for a
variable star formation efficiency
in protoclusters is itself a new target for theoretical attack, most likely
through a general calculation of star formation and feedback such as that
described at the beginning of this Section. As was also mentioned there, if
such models can be made to agree with the data in Fig.~\ref{fig6}, they will
likely also shed new light on the empirical efficiency
of cluster formation, and perhaps on other generic properties of GCSs as
well---and, thence, on larger-scale processes in galaxy formation.

\begin{acknowledgements}

This work was supported by NASA through grant number HF-1097.01-97A awarded by
the Space Telescope Science Institute, which is operated by the Association of
Universities for Research in Astronomy, Inc., for NASA under contract
NAS5-26555.

\end{acknowledgements}


\begin{thebibliography}{}

\bibitem[\protect\astroncite{Aguilar et al.}{1988}]{agu88}
Aguilar, L., Hut, P., \& Ostriker, J. P. 1988, ApJ, 335, 720

\bibitem[\protect\astroncite{Ashman \& Zepf}{1992}]{ash92}
Ashman, K. M., \& Zepf, S. E. 1992, ApJ, 384, 50

\bibitem[\protect\astroncite{Ashman \& Zepf}{1998}]{ash98}
Ashman, K. M., \& Zepf, S. E. 1998, Globular Cluster Systems (New York:
Cambridge Univ.~Press)

\bibitem[\protect\astroncite{Bellazzini}{1998}]{bel98}
Bellazzini, M. 1998, New Astronomy, 3, 219

\bibitem[\protect\astroncite{Bellazzini et al.}{1996}]{bel96}
Bellazzini, M., Vesperini, E., Ferraro, F. R., \& Fusi Pecci, F. 1996,
MNRAS, 279, 337

\bibitem[\protect\astroncite{Bender et al.}{1992}]{ben92}
Bender, R., Burstein, D., \& Faber, S. 1992, ApJ, 399, 462

\bibitem[\protect\astroncite{Blakeslee}{1997}]{bla97}
Blakeslee, J. P. 1997, ApJ, 481, L59

\bibitem[\protect\astroncite{Blakeslee}{1999}]{bla99}
Blakeslee, J. P. 1999, AJ, 118, 1506

\bibitem[\protect\astroncite{Blakeslee et al.}{1997}]{btm97}
Blakeslee, J. P., Tonry, J. L., \& Metzger, M. R. 1997, AJ, 114, 482

\bibitem[\protect\astroncite{Caputo \& Castellani}{1984}]{cap84}
Caputo, F., \& Castellani, V. 1984, MNRAS, 207, 185

\bibitem[\protect\astroncite{Capuzzo-Dolcetta}{1993}]{cap93}
Capuzzo-Dolcetta, R. 1993, ApJ, 415, 616

\bibitem[\protect\astroncite{C\^ot\'e et al.}{1998}]{cot98}
C\^ot\'e, P., Marzke, R. O., \& West, M. J. 1998, ApJ, 501, 554

\bibitem[\protect\astroncite{C\^ot\'e et al.}{2000}]{cot00}
C\^ot\'e, P., Marzke, R. O., West, M. J., \& Minniti, D. 2000, ApJ, in press
({\tt astro-ph/9911527})

\bibitem[\protect\astroncite{de Vaucouleurs \& Nieto}{1978}]{dvn78}
de Vaucouleurs, G., \& Nieto, J.-L. 1978, ApJ, 220, 449

\bibitem[\protect\astroncite{Dekel \& Silk}{1986}]{dek86}
Dekel, A.., \& Silk, J. 1986, ApJ, 303, 39

\bibitem[\protect\astroncite{Djorgovski}{1995}]{djo95}
Djorgovski, S. 1995, ApJ, 438, L29

\bibitem[\protect\astroncite{Djorgovski \& Meylan}{1994}]{djo94}
Djorgovski, S., \& Meylan, G. 1994, AJ, 108, 1292

\bibitem[\protect\astroncite{Durrell et al.}{1996}]{dur96}
Durrell, P. R., Harris, W. E., Geisler, D., \& Pudritz, R. E. 1996,
AJ, 112, 972

\bibitem[\protect\astroncite{Elmegreen}{2000}]{elm00}
Elmegreen, B. G. 2000, in Toward a New Millennium in Galaxy Morphology, ed.
D. L. Block, I. Puerari, A. Stockton, and D. Ferreira (Dordrecht: Kluwer),
in press ({\tt astro-ph/9911157})

\bibitem[\protect\astroncite{Elmegreen \& Clemens}{1985}]{elm85}
Elmegreen, B. G., \& Clemens, C. 1985, ApJ, 294, 523

\bibitem[\protect\astroncite{Elmegreen \& Efremov}{1997}]{elm97}
Elmegreen, B. G., \& Efremov, Y. N. 1997, ApJ, 480, 235

\bibitem[\protect\astroncite{Fall \& Rees}{1977}]{fal77}
Fall, S. M., \& Rees, M. J. 1977, MNRAS, 181, 37P

\bibitem[\protect\astroncite{Fall \& Rees}{1985}]{fal85}
Fall, S. M., \& Rees, M. J. 1985, ApJ, 298, 18

\bibitem[\protect\astroncite{Forbes et al.}{1997}]{for97}
Forbes, D. A., Brodie, J. P., \& Grillmair, C. J. 1997, AJ, 113, 1652

\bibitem[\protect\astroncite{Gnedin \& Ostriker}{1997}]{gne97}
Gnedin, O. Y., \& Ostriker, J. P. 1997, ApJ, 474, 223

\bibitem[\protect\astroncite{Goodwin}{1997}]{goo97}
Goodwin, S. P. 1997, MNRAS, 284, 785

\bibitem[\protect\astroncite{Harris}{1986}]{har86}
Harris, W. E. 1986, AJ, 91, 822

\bibitem[\protect\astroncite{Harris}{1991}]{har91}
Harris, W. E. 1991, ARA\&A, 29, 543

\bibitem[\protect\astroncite{Harris}{1996}]{har96}
Harris, W. E. 1996, AJ, 112, 1487

\bibitem[\protect\astroncite{Harris}{2000}]{har00}
Harris, W. E. 2000, Globular Cluster Systems: Lectures for the 1998 Saas-Fee
Advanced Course on Star Clusters (Swiss Society for Astrophysics and
Astronomy), in press (preprint at
{\tt http://physun.mcmaster.ca/$\sim$harris/WEHarris.html})

\bibitem[\protect\astroncite{Harris \& Pudritz}{1994}]{har94}
Harris, W. E., \& Pudritz, R. E. 1994, ApJ, 429, 177

\bibitem[\protect\astroncite{Harris \& van den Bergh}{1981}]{hvd81}
Harris, W. E., \& van den Bergh, S. 1981, AJ, 86, 1627

\bibitem[\protect\astroncite{Harris et al.}{1998}]{har98}
Harris, W. E., Harris, G. L. H., \& McLaughlin, D. E. 1998, AJ, 115, 1801

\bibitem[\protect\astroncite{Hills}{1980}]{hil80}
Hills, J. G. 1980, ApJ, 225, 986

\bibitem[\protect\astroncite{King}{1966}]{kin66}
King, I. R. 1966, AJ, 71, 64

\bibitem[\protect\astroncite{Kissler-Patig}{1997}]{kis97}
Kissler-Patig, M. 1997, A\&A, 319, 83

\bibitem[\protect\astroncite{Kissler-Patig \& Gebhardt}{1999}]{kig99}
Kissler-Patig, M., \& Gebhardt, K. 1999, AJ, 118, 1526

\bibitem[\protect\astroncite{Kissler-Patig et al.}{1998}]{kis98}
Kissler-Patig, M., Forbes, D. A., \& Minniti, D. 1998, MNRAS, 298, 1123

\bibitem[\protect\astroncite{Lada et al.}{1991}]{lad91}
Lada, E. A., DePoy, D. L., Evans, N. J., \& Gatley, I. 1991, ApJ, 371, 171

\bibitem[\protect\astroncite{Lada}{1992}]{lad92}
Lada, E. A. 1992, ApJ, 393, L25

\bibitem[\protect\astroncite{Larson}{1988}]{lar88}
Larson, R. B. 1988, in IAU Symposium No. 126, Globular Cluster Systems in
Galaxies, ed. J. E. Grindlay and A. G. D. Philip (Dordrecht: Kluwer), 311

\bibitem[\protect\astroncite{Larson}{1993}]{lar93}
Larson, R. B. 1993, in ASP Conf. Ser. 48, The Globular Cluster--Galaxy
Connection, ed. G. H. Smith and J. P. Brodie (San Francisco: ASP), 675

\bibitem[\protect\astroncite{Mac\,Low \& Ferrara}{1999}]{mac99}
Mac\,Low, M.-M., \& Ferrara, A. 1999, ApJ, 513, 142

\bibitem[\protect\astroncite{Mandushev et al.}{1991}]{man91}
Mandushev, G., Spassova, N., \& Staneva, A. 1991, A\&A, 252, 94

\bibitem[\protect\astroncite{Mathieu}{1983}]{mat83}
Mathieu, R. D. 1983, ApJ, 267, L97

\bibitem[\protect\astroncite{McLaughlin}{1995}]{mcl95}
McLaughlin, D. E. 1995, AJ, 109, 2034

\bibitem[\protect\astroncite{McLaughlin}{1999}]{mcl99}
McLaughlin, D. E. 1999, AJ, 117, 2398

\bibitem[\protect\astroncite{McLaughlin}{2000}]{mcl00}
McLaughlin, D. E. 2000, ApJ, in press

\bibitem[\protect\astroncite{McLaughlin \& Pudritz}{1996}]{mcl96}
McLaughlin, D. E., \& Pudritz, R. E. 1996, ApJ, 457, 578

\bibitem[\protect\astroncite{McLaughlin et al.}{1993}]{mcl93}
McLaughlin, D. E., Harris, W. E., \& Hanes, D. A. 1993, ApJ, 409, L45

\bibitem[\protect\astroncite{Meurer et al.}{1995}]{meu95}
Meurer, G. R., Heckman, T. M., Leitherer, C., Kinney, A., Robert, C.,
\& Garnett, D. R. 1995, AJ, 110, 2665

\bibitem[\protect\astroncite{Meylan \& Heggie}{1997}]{mey97}
Meylan, G., \& Heggie, D. C. 1997, A\&A Rev., 8, 1

\bibitem[\protect\astroncite{Miller et al.}{1998}]{mil98}
Miller, B. W., Lotz, J. M., Ferguson, H. C., Stiavelli, M., \& Whitmore,
B. C. 1998, ApJ, 508, L133

\bibitem[\protect\astroncite{Murali \& Weinberg}{1997a}]{mwa97}
Murali, C., \& Weinberg, M. D. 1997a, MNRAS, 288, 749

\bibitem[\protect\astroncite{Murali \& Weinberg}{1997b}]{mwb97}
Murali, C., \& Weinberg, M. D. 1997b, MNRAS, 288, 767

\bibitem[\protect\astroncite{Murali \& Weinberg}{1997c}]{mwc97}
Murali, C., \& Weinberg, M. D. 1997c, MNRAS, 291, 717

\bibitem[\protect\astroncite{Murray \& Lin}{1992}]{mur92}
Murray, S. D., \& Lin, D. N. C. 1992, ApJ, 400, 265

\bibitem[\protect\astroncite{Murray \& Lin}{1993}]{mur93}
Murray, S. D., \& Lin, D. N. C. 1993, in The Globular Cluster--Galaxy
Connection, ed. G. H. Smith and J. P. Brodie (San Francisco: ASP), 738

\bibitem[\protect\astroncite{Ostriker et al.}{1989}]{ost89}
Ostriker, J. P., Binney, J., \& Saha, P. 1989, MNRAS, 241, 849

\bibitem[\protect\astroncite{Patel \& Pudritz}{1994}]{pat94}
Patel, K., \& Pudritz, R. E. 1994, ApJ, 424, 688

\bibitem[\protect\astroncite{Peebles \& Dicke}{1968}]{pee68}
Peebles, P. J. E., \& Dicke, R. H. 1968, ApJ, 154, 891

\bibitem[\protect\astroncite{Pryor \& Meylan}{1993}]{pry93}
Pryor, C., \& Meylan, G. 1993, in ASP Conf. Ser. 50, Structure and Dynamics
of Globular Clusters, ed. S. G. Djorgovski and G. Meylan (San Francisco: ASP),
357

\bibitem[\protect\astroncite{Saito}{1979a}]{sai79a}
Saito, M. 1979a, PASJ, 31, 181

\bibitem[\protect\astroncite{Saito}{1979b}]{sai79b}
Saito, M. 1979b, PASJ, 31, 193

\bibitem[\protect\astroncite{Schweizer}{1987}]{sch87}
Schweizer, F. 1987, in Nearly Normal Galaxies, ed. S. M. Faber
(New York: Springer), 18

\bibitem[\protect\astroncite{Schweizer}{1999}]{sch99}
Schweizer, F. 1999, in Spectrophotometric Dating of Stars and Galaxies, ed.
I. Hubeny, S. Heap, and R. Cornett (San Francisco: ASP), in press
({\tt astro-ph/9906488})

\bibitem[\protect\astroncite{Searle \& Zinn}{1978}]{sea78}
Searle, L., \& Zinn, R. 1978, ApJ, 225, 357

\bibitem[\protect\astroncite{Spitzer}{1987}]{spi87}
Spitzer, L. 1987, Dynamical Evolution of Globular Clusters (Princeton:
Princeton Univ.~Press)

\bibitem[\protect\astroncite{van den Bergh}{1994}]{vdb94}
van den Bergh, S. 1994, ApJ, 435, 203

\bibitem[\protect\astroncite{van der Marel}{1991}]{vdm91}
van der Marel, R. P. 1991, MNRAS, 253, 710

\bibitem[\protect\astroncite{Verschueren}{1990}]{ver90}
Verschueren, W. 1990, A\&A, 234, 156

\bibitem[\protect\astroncite{Vesperini}{1997}]{ves97}
Vesperini, E. 1997, MNRAS, 287, 915

\bibitem[\protect\astroncite{West et al.}{1995}]{wes95}
West, M. J., C\^ot\'e, P., Jones, C., Forman, W., \& Marzke, R. O. 1995,
ApJ, 453, L77

\bibitem[\protect\astroncite{Whitmore \& Schweizer}{1995}]{whi95}
Whitmore, B. C., \& Schweizer, F. 1995, AJ, 109, 960

\bibitem[\protect\astroncite{Zepf \& Ashman}{1993}]{zep93}
Zepf, S. E., \& Ashman, K. M. 1993, MNRAS, 264, 611

\bibitem[\protect\astroncite{Zepf et al.}{1999}]{zep99}
Zepf, S. E., Ashman, K. M., English, J., Freeman, K. C., \& Sharples, R. M.
1999, AJ, 118, 752

\end{thebibliography}
\end{document}